\documentclass[12pt]{article}
\usepackage{amsfonts,amssymb,epsfig}
\usepackage[latin1]{inputenc}
\usepackage[english]{babel}
\usepackage{hyperref}
\usepackage{color}
\newtheorem{thm}{Th\'eor\`eme}[section]
\newtheorem{cor}[thm]{Corollaire}
\newtheorem{lem}[thm]{Lemme}
\newtheorem{pro}[thm]{Proposition}
\newtheorem{dfn}[thm]{D\'efinition}
\newtheorem{rmk}[thm]{Remark}
\newtheorem{expl}[thm]{Exemple}

\def\dessous#1\sous#2{\mathrel{\mathop{\kern0pt#2}\limits_{#1}}}

\newcommand{\beq}{\begin{eqnarray}}
\newcommand{\eeq}{\end{eqnarray}}
\newcommand{\bpro}{\begin{pro}}
\newcommand{\epro}{\end{pro}}
\newcommand{\blem}{\begin{lem}}
\newcommand{\elem}{\end{lem}}
\newcommand{\bdfn}{\begin{dfn}}
\newcommand{\edfn}{\end{dfn}}
\newcommand{\bcor}{\begin{cor}}
\newcommand{\ecor}{\end{cor}}
\newcommand{\bthm}{\begin{thm}}
\newcommand{\ethm}{\end{thm}}
\newcommand{\bex}{\begin{expl}}
\newcommand{\eex}{\end{expl}}
\newcommand{\brmk}{\begin{rmk}}
\newcommand{\ermk}{\end{rmk}}
\newcommand{\benum}{\begin{enumerate}}
\newcommand{\eenum}{\end{enumerate}}
\newcommand{\bitem}{\begin{itemize}}
\newcommand{\eitem}{\end{itemize}}

\begin{document}
\begin{center}
{\Large\bf {Teleportation of a qubit using exotic entangled coherent states }}

 \vspace{0.5cm}

 Isiaka Aremua$^{1}$ and Laure Gouba$^{2}$ 

 $^{1}${\em
Universit\'{e} de Lom\'{e} (UL), Facult\'{e} Des Sciences  (FDS), D\'{e}partement de Physique \\
{ Laboratoire de Physique des Mat\'eriaux et des Composants \`a Semi-Conducteurs}\\
Centre d'Excellence R\'egional pour la Ma\^itrise de l'Electricit\'e (CERME)\\
{ Universit\'{e} de Lom\'{e} (UL), 01 B.P. 1515  Lom\'{e} 01, Togo.}
E-mail: claudisak@gmail.com}\\
 $^{2}${\em
The Abdus Salam International Centre for Theoretical Physics (ICTP),
Strada Costiera 11, I-34151 Trieste Italy. E-mail: laure.gouba@gmail.com}

\vspace{1.0cm}

\today

\begin{abstract}
\noindent
In this paper, we study the exotic Landau problem at the classical level where two conserved quantities are derived. At the quantum level, the corresponding quantum operators of the conserved quantities provide two oscillator representations from which we derive two Boson Fock spaces. Using the normalized coherent states which are the minimum uncertainty states on noncommutative configuration space isomorphic to each of the boson Fock space, we form entangled coherent states which are Bell- like states labeled quasi-Bell states. The effect of non-maximality of a quasi-Bell state based quantum channel is investigated in the context of a teleportation of a qubit. 
\end{abstract}

\end{center}

\setcounter{footnote}{0}

\section{Introduction}
 
 We have recently been interested in the study of a system of an electron moving on a plane in uniform external magnetic and electric fields where we constructed different classes of coherent states in the context of discrete and continuous spectra, and the situation where both spectra are purely discrete \cite{aremua-gouba1, aremua-gouba2}. Following these works, we investigated the action of unitary maps on the associated quantum Hamiltonians and constructed the coherent states of the Gazeau-Klauder type \cite{aremua-gouba}. The idea of the construction of these coherent states follows previous methods from the literature \cite{gazeauklauder, alibagarello, gazeau-novaes, lgouba}.
 
In this paper, we consider the motion of charged particles in a flat noncommutative plane $xy$, with a constant magnetic field applied along the z-axis, referring to the exotic Landau problem 
\cite{peter1, peter2, peter3, peter4, aremua-hounkonnou-balo}.  Two commuting conserved quantities are derived from the study of the exotic model at the classical level in the situation of a pure magnetic case. Through  canonical quantization, the classical quantities are promoted to operators labeled by " hats " and the Poisson brackets are replaced by $(i/\hbar)$ multiplied by the commutators. Using two conserved quantities 
operators, two oscillator representations allow to have an explicit form of the wave functions of the quantum Hilbert space.

The two oscillators systems generated by the two conserved quantities are labeled "system A" and "system B". For each of the systems, we determine the minimum uncertainty coherent states on noncommutative configuration space isomorphic to the boson Fock space. Using these coherent states we form entangled coherent states in the sense of the works done by Sanders et al. \cite{sanders1, sanders2, sanders3, sanders4} from which we form the quasi- Bell states.
Following the teleportation protocol by Bennett et al. \cite{bennett}, we perform a  teleportation of a qubit using one of the quasi-Bell states as a channel has been investigated and the minimum assured fidelity (MASFI) by this channel is determined followed by the computation of the fidelity of sending a qubit.

The paper is organized as follows. Section \ref{sec2} is about  the exotic Landau problem presented at the classical level as well as at the quantum level. In section \ref{sec3}, the exotic quasi-Bell states are presented. The section \ref{sec4} is dedicated to the teleportation of a qubit and the fidelity of teleportation. Some concluding remarks are given in section \ref{sec5}.

\section{The exotic Landau problem}\label{sec2}

This section is strongly derived from previous works \cite{ peter1, peter2, peter3, peter4,aremua-hounkonnou-balo} where the details of the calculations and more explanations  are given.
{\subsection{The model at the classical level}
We consider the two-dimensional noncommutative plane where the  fundamental commutation relations are given by 
\begin{equation}\label{fcr1}
\{x_i, x_j \} = \theta \varepsilon^{ij}; \quad 
\{x_i, p_j\} = \delta^{ij}; \quad 
\{ p_i, p_j\} = 0,
\end{equation}
with
$\varepsilon^{ij}$ are the components of the
antisymmetric tensor normalized by $\varepsilon^{12} =1$, $\delta_{ij}$ is the Kroenecker delta and $\theta $ is the noncommutative parameter. 
The associated Poisson bracket on phase space differ from the canonical one by an additional term as follows
\begin{equation}
\{f, g \} = \frac{\partial f}{\partial \vec{x}}\cdot 
\frac{\partial g}{\partial \vec{p}} 
-\frac{\partial g}{\partial \vec x}\frac{\partial f}{\partial \vec p} 
+\theta \left( \frac{\partial f}{\partial x_1}\frac{\partial g}{\partial x_2} - \frac{\partial g}{\partial x_1}\frac{\partial f}{\partial x_2}
\right).
\end{equation}
For a system of one charged particle of mass $M$ and charge e moving in this plane, we choose the noncommutative parameter $\theta$ to be exotic in the sense that we relate it  to the  \textquotedblleft exotic \textquotedblright parameter 
$\kappa$ as follows
\begin{equation}
\theta  = \frac{\kappa}{M^2},
\end{equation} 
and the system  of this exotic particle is described by the Hamiltonian 
\beq{\label{sol011}}
 \mathcal H = \frac{1}{2M}\sum_{i = 1}^2 p_i^{2} + e V(x_1,x_2),\quad i = 1, 2, 
\eeq
where $V$ is the electric potential, assumed to be time independent.

In the presence of an electromagnetic field, where the electric field $\vec E$ and the magnetic field $\vec B$ are assumed constant, the Hamiltonian remains the standard one in (\ref{sol011}) while the Poisson bracket is modified to 
\begin{equation}
\{f, g \} = \frac{\partial f}{\partial \vec{x}}\cdot 
\frac{\partial g}{\partial \vec{p}} 
-\frac{\partial g}{\partial \vec x}\frac{\partial f}{\partial \vec p} 
+\theta \left( \frac{\partial f}{\partial x_1}\frac{\partial g}{\partial x_2} - \frac{\partial g}{\partial x_1}\frac{\partial f}{\partial x_2}
\right) + B \left( \frac{\partial f}{\partial p_1}\frac{\partial g}{\partial p_2} - \frac{\partial g}{\partial p_1}\frac{\partial f}{\partial p_2}
\right).
\end{equation}
The fundamental commutations relations (\ref{fcr1}) are now
\beq\label{pbrack}
\{x_{i}, x_{j}\}  = \frac{M}{M^*}\theta \varepsilon^{ij}, 
\qquad \{x_{i}, p_{j}\}  = \frac{M}{M^*}\delta^{ij}, \qquad \{p_{i}, p_{j}\} = \frac{M}{M^*}eB\varepsilon^{ij},
\eeq
where the noncommutative parameter $\theta$ and the charge $e$ combine with the magnetic field $B$ into an effective mass given by 
$M^* = M(1 - e\theta B)$
THe vector potential is chosen as 
 $ A_i = \frac{1}{2} B\epsilon_{ij}x_j $ and the electric field $E_i = -\partial_i V$.
The equations of motion are obtained through the relations 
$\dot{\chi} = \{ H , \chi \} $, with $\chi  \in (x_1, x_2, p_1, p_2)\:\:, i = 1,2$
\beq\label{equa00}
M^{*}\dot{x}_{i} = p_{i} - M e \theta\varepsilon^{ij}E^{j}, \qquad \dot{p}_{i} = e B \varepsilon^{ij}\dot{x}_{j} + e E^{i}, \:\: i, j = 1,2.
\eeq 

Let's consider the situation of a pure magnetic case, means $ E= 0 $. Then the particle performs the usual cyclotronic motion but with modified frequency 
$\omega^* = \frac{\omega}{1 -e\theta B}$, that is 
\begin{equation}
x_i(t) = R(- \omega^* t) \alpha_i + \beta_i
\end{equation}
where $\vec{\alpha} = (\alpha_1,\alpha_2)$ and $\vec{\beta} = (\beta_1,\beta_2)$ are constant vectors.

The time-dependent translation or ``boost" 
\begin{equation}
x_i \rightarrow x_i +b_i; \quad p_i \rightarrow p_i + M^*\dot{b}_i
\end{equation}
is a symmetry for the equation (\ref{equa00}) whenever $\vec{b} = (b_1,\;b_2)$ satisfies 
\begin{equation}\label{equa01}
M^* \ddot{b}_i - eB\varepsilon^{ij}\dot{b}_j = 0,
\end{equation}
and the equation (\ref{equa01}) is solved as 
\begin{equation}
b_i(t) = R(-\omega^* t) a_i + c_i,
\end{equation}
where $\vec{a} = (a_1, a_2)$ and $\vec{c} = (c_1, c_2)$ are constant vectors. The associated conserved quantities are therefore
\begin{equation}
\mathcal{P}_i = M^* (\dot{x}_i - \omega^*\varepsilon^{ij} x_j); \quad 
\mathcal{K}_i = \frac{M^*}{M}R(\omega^* t)p_i = \frac{{M^*}^2}{M}R(\omega^* t)\dot{x}_i,\quad i=1,2.
\end{equation}
which follows the following algebra
\begin{equation}
\{\mathcal{P}_i,\: \mathcal{P}_j\} = -M^*\omega^*\varepsilon^{ij};\quad 
\{\mathcal{K}_i,\;\mathcal{K}_j\} = (1 -e\theta B)M^*\omega^*\varepsilon^{ij};\quad 
\{\mathcal{P}_i, \quad \mathcal{K}_j\} = 0. 
\end{equation}

\subsection{Model at the quantum level}

At the quantum level, the classical quantities are promoted to operators labeled with `` hats"  and the Poisson brackets are replaced by commutators multiplied by the factor $i\hbar$. Due to the exotic noncommutator parameter, the position representation cannot be performed here.

Still in the condition where $E = 0$, $e B\theta \neq 1$, 
the quantum Hamiltonian 
\begin{equation}\label{jjjj}
\hat H = \sum_{i = 1}^2\frac{\hat {p_i}^2}{2M},\quad i = 1,2, 
\end{equation}
depends only on the conserved quantities 
$\hat{\mathcal{K}}_i,  \: i = 1, 2$
 which satisfies the following commutations relations
\begin{equation} 
[\hat{\mathcal{K}}_i,\;\hat{\mathcal{K}}_j]  = i\hbar(1 -e\theta B) M^* \omega^* \varepsilon^{ij}. 
\end{equation}
 We define then the operators 
$a, a^\dagger$ such that 
\begin{equation}\label{ladder00}
\hat a =    \hat{\mathcal K}^{1} + i\hat{\mathcal K}^{2}, \qquad  \hat a^{\dag} =   
\hat{\mathcal K}^{1} - i   \hat{\mathcal K}^{2}, \quad[\hat a, \hat a^{\dag}] = 2\hbar (1-eB\theta) M \omega.
\end{equation}
The quantum hamiltonian  becomes (\ref{jjjj}) becomes
\begin{equation}\label{operat01}
\hat{H} = \frac{1}{2M(1-eB\theta)^{2}}\hat a^{\dag}\hat a + \frac{ \hbar \omega^{*}}{2}\,,
\end{equation}
where   $\omega^{*} = {eB}/{M^{*}},  M^{*} = (1-eB\theta)M$. It is convenient to define the  creation and annihilator operators
 $\{\mathfrak a,  {\mathfrak a}^{\dag}\}$  as follows
 
  \begin{equation}\label{algeb00}
 \mathfrak a 
 = \frac{1}{ \sqrt{2\hbar (1-eB\theta)M\omega}}\, \hat a \quad 
{\mathfrak a}^{\dag} =  \frac{1}{\sqrt{2\hbar (1-eB\theta)M\omega}} \hat a^{\dag} 
\end{equation}

that satisfy the Fock algebra $
 [\mathfrak a,  {\mathfrak a}^{\dag}] = \mathbb {I}$. The noncommutative configuration space in this sector is then isomorphic to the boson Fock space 
 
 \begin{equation}
 \Gamma_{\mathcal K} = \textrm{span} \left\lbrace 
 | n\rangle \equiv \frac{1}{\sqrt{n!}}({\mathfrak{a}^\dagger})^n | 0 \rangle_{\mathcal K}\right\rbrace_{n= 0}^\infty.
 \end{equation}
 
Let's consider now the oscillator representation of the other conserved quantity, $\mathcal{P}_i, \: i = 1,2$, 
which are  \textquotedblleft ${\hat{x}_i, i = 1,2 }$-only operators \textquotedblright, as follows
\begin{equation}
\hat b = \hat{\mathcal P}^{1} + i \hat{\mathcal P}^{2}, \quad  
\hat b^{\dag} = \hat{\mathcal P}^{1} - i \hat{\mathcal P}^{2}, \qquad  [\hat b, \hat b^{\dag}] = 2\hbar M\omega\,.
\end{equation}

In the same manner as above, it is convenient to introduce the operators 
$\{\mathfrak b, \mathfrak b^{\dag}\}$
\begin{equation}
 \mathfrak b = \frac{1}{\sqrt{2\hbar M \omega}}\, \hat b, \quad 
\mathfrak b^{\dag} =  \frac{1}{\sqrt{2\hbar M \omega }}\hat b^{\dag},
\end{equation}
that satisfy the Fock algebra $ [\mathfrak b, \mathfrak b^{\dag}] = \mathbb I $. The non-commutative configuration in this sector is then isomorphic to the boson Fock space 
\begin{equation}
 \Gamma_{\mathcal P} = \textrm{span} \left\lbrace 
 | m\rangle \equiv \frac{1}{\sqrt{n!}}({\mathfrak{b}^\dagger})^m | 0 \rangle_{\mathcal P}\right\rbrace_{m= 0}^\infty.
 \end{equation}
 
Let's consider now the boson Fock space of the system  as
$\Gamma = \Gamma_{\mathcal P} \otimes \Gamma_{\mathcal K}$ such that 
\begin{equation}
\Gamma = \textrm{span}\left\lbrace |m \rangle \otimes |n \rangle = |m, n\rangle \equiv \frac{1}{\sqrt{m !n !}}(\mathfrak {b}^{\dag})^{m}(\mathfrak {a}^{\dag})^{n}|0,0\rangle_{\mathcal{K}, \mathcal{P}}
\right\rbrace^{\infty}_{m,n =0}\,.
\end{equation}

The energy only depends on the $\mathcal{K}_i,\: i = 1,2 $-dynamics as the second-oscillator operators have no contribution. The energy levels are then  
\begin{equation}\label{eigval00}
E_{ n} = \hbar \omega^{*}\left(n + \frac{1}{2}\right).
\end{equation} 

The wave function of the quantum Hilbert space are given by 
$|\Psi \rangle = |n,m \rangle$.
 
\section{Exotic quasi-Bell states}\label{sec3}
 
 The minimal uncertainty states on noncommutative configuration space, which is isomorphic to the boson Fock space $\Gamma_{\mathcal K}$ are well-known to be the normalized coherent states 
\begin{equation}\label{estate}
| \alpha\rangle =  e^{-\alpha \bar{\alpha}/2} e^{\alpha \mathfrak{a}^\dagger}| 0 \rangle_{\mathcal{K}},
\end{equation}
where $\alpha$ is a dimensionless complex number. These states provide an overcomplete basis on the non-commutative configuration space.
In the same spirit, the minimal uncertainty states on noncommutative configuration space, which is isomorphic to the boson Fock space $\Gamma_{\mathcal P}$ are well-known to be the normalized coherent states 
\begin{equation}\label{state}
| \beta \rangle =  e^{-\beta \bar{\beta}/2} e^{\beta \mathfrak{b}^\dagger}| 0 \rangle_{\mathcal{P}},
\end{equation}
where $\beta$ is a dimensionless complex number.

One can notice that the state $|\alpha \rangle $ in equation (\ref{estate}) depend implicitly on the exotic noncommutative parameter $\theta$ while it is not the case for the states $| \beta \rangle$ in the equation (\ref{state}), and that $\langle \alpha \vert \beta \rangle = 0.$

We consider now two coherent states $\{ | \alpha \rangle, |-\alpha \rangle \}$  of the first oscillator system A,  which satisfy $\langle \alpha | -\alpha \rangle = \exp\{-2|\alpha |^2\}$ meaning their non-orthogonality. In the same spirit, we consider two coherent states $\{ |\beta \rangle, |-\beta \rangle \}$ of the second oscillator system B, which satisfy $\langle \beta | - \beta \rangle = \exp\{-2|\beta |^2\}$

Let us now construct the following normalized states 
\begin{eqnarray}\label{qb1}
|\Psi_1 \rangle_{\alpha, \beta} & = & \frac{1}{\sqrt{2(1 + e^{-2 (|\alpha|^2 + |\beta|^2)})}} \left( 
|\alpha \rangle |-\beta\rangle + |-\alpha\rangle|\beta\rangle \right)\\\label{qb2}
|\Psi_2 \rangle_{\alpha, \beta} &= & 
\frac{1}{\sqrt{2(1 - e^{-2 (|\alpha|^2 + |\beta|^2)})}}
 \left( 
|\alpha \rangle |-\beta\rangle - |-\alpha\rangle|\beta\rangle \right) \\\label{qb3}
|\Psi_3 \rangle_{\alpha, \beta} &= & \frac{1}{\sqrt{2(1 + e^{-2 (|\alpha|^2 + |\beta|^2)})}} \left( 
|\alpha \rangle |\beta\rangle + |-\alpha\rangle|-\beta\rangle \right)\\\label{qb4}
|\Psi_4 \rangle_{\alpha, \beta} &= & \frac{1}{\sqrt{2(1 - e^{-2 (|\alpha|^2 + |\beta|^2)})}} \left( 
|\alpha \rangle | \beta\rangle - |-\alpha\rangle|-\beta\rangle \right)
\end{eqnarray}
The states $| \Psi_i \rangle_{\alpha,\beta}, \: i = 1,2,3,4$ are not orthogonal to each other and this is justify by their Gram matrix 
\begin{equation}
G_{ij} = |{}_{\alpha,\beta\langle\, \Psi_i\,|\, \Psi_j\,\rangle_{\alpha,\beta}}|,\:\: i =1 \ldots 4,\: j = 1 \ldots 4,
\end{equation} 
which is 
\begin{equation}
G = \left( 
\begin{array}{cccc}
1 & 0 & G_{1,3} & 0 \\
0 & 1 & 0 & G_{2,4} \\
G_{3,1} & 0 & 1 & 0 \\
0 & G_{4,2} & 0 & 1
\end{array}
\right)\,,
\end{equation}
with $G_{1,3} = G_{3,1} = \frac{e^{-2|\alpha|^2} + e^{-2|\beta|^2}}{1 + e^{-2(|\alpha|^2 + |\beta|^2)}}$ and 
$G_{2,4} = G_{4,2} = \frac{e^{-2|\alpha|^2} - e^{-2|\beta|^2}}{1 - e^{-2(|\alpha|^2 + |\beta|^2)}}$.  
In order to determine whether these non-orthogonal states are entangled or not, we study here the entropy of entanglement as described in \cite{gnonfin}. Let's denote by $\rho^i_{\alpha, \beta}, \: i = 1,\ldots 4$, the density operators of these states as follows
\begin{equation}
\rho^{(i)}_{\alpha,\beta} = |\Psi_i\rangle_{\alpha,\beta}{}_{\alpha,\beta}\langle \Psi_i |.
\end{equation}
The reduced density operators are $\rho_\alpha^{(1)} = \rho_\alpha^{(3)}$ and $\rho_\alpha^{(2)} = \rho_\alpha^{(4)}$, 
with 
\begin{eqnarray}\nonumber
\rho_\alpha^{(1)} &=& \frac{1}{2( 1+ e^{-2(|\alpha|^2 + |\beta|^2})}\\\nonumber
&\times&
\left( 
(1 + e^{-4|\beta|^2})|\alpha\rangle\langle \alpha |
+ 2 e^{-2|\beta|^2}| \alpha \rangle\langle -\alpha |
+ 2e^{-2|\beta|^2 }|-\alpha \rangle \langle \alpha |
+ (1 + e^{-4|\beta|^2})|-\alpha\rangle\langle -\alpha |
\right)\\
{}
\end{eqnarray}
\begin{eqnarray}\nonumber
\rho_\alpha^{(2)} &=& \frac{1}{2( 1- e^{-2(|\alpha|^2 + |\beta|^2})}\\\nonumber
&\times&
\left( 
(1 + e^{-4|\beta|^2})|\alpha\rangle\langle \alpha |
- 2 e^{-2|\beta|^2}| \alpha \rangle\langle -\alpha |
- 2e^{-2|\beta|^2 }|-\alpha \rangle \langle \alpha |
+ (1 + e^{-4|\beta|^2})|-\alpha\rangle\langle -\alpha |
\right)\\
{}
\end{eqnarray}
The eigenvalues of the density operators $\rho_\alpha^{(1)}$  or $\rho_\alpha^{(3)}$ are the following 
\begin{equation}
\lambda_1 = \frac{(1 - e^{-2|\beta|^2})^2}{2(1 + e^{-2(|\alpha|^2 + |\beta|^2)})};\: 
\lambda_1' = \frac{(1 + e^{-2|\beta|^2})^2}{2(1 + e^{-2(|\alpha|^2 + |\beta|^2)})}
\end{equation}
and for $\rho_\alpha^{(2)}$  or $\rho_\alpha^{(4)}$, they are 
\begin{equation}
\lambda_2 = \frac{(1 - e^{-2|\beta|^2})^2}{2(1 - e^{-2(|\alpha|^2 + |\beta|^2)})};\: 
\lambda_2' = \frac{(1 + e^{-2|\beta|^2})^2}{2(1 - e^{-2(|\alpha|^2 + |\beta|^2)})}
\end{equation}
Hence the entropy of entanglement is 
\begin{equation}
E(|\Psi_1\rangle_{\alpha,\beta}) = E (\Psi_3\rangle_{\alpha, \beta}) = -\lambda_1\log \lambda_1
-\lambda_1'\log \lambda_1'
\end{equation}
and 
\begin{equation}
E(|\Psi_2\rangle_{\alpha,\beta}) = E (\Psi_4\rangle_{\alpha, \beta}) = -\lambda_2\log \lambda_2
-\lambda_2'\log \lambda_2'
\end{equation}
These states in (\ref{qb1}), (\ref{qb2}), (\ref{qb3}), (\ref{qb4}) are entangled and in the limits $|\alpha| \rightarrow +\infty $ and $|\beta| \rightarrow + \infty$ they are maximally entangled states. These entangled coherent states are quasi-Bell states and the dimension of the space spanned by these states is 4 even though they are embedded in a vector space of infinite dimension.

\section{Quantum teleportation of a qubit}\label{sec4}

In this section, we formulate the teleportation protocole between two friends Amy and Bella as described in \cite{bennett, popescu, horodecki}. We assume that Amy and Bella are far away and sharing the quantum channel
\begin{equation}\label{abq}
|\Psi_3 \rangle_{\alpha, \beta} =  \frac{1}{\sqrt{2(1 + e^{-2 (|\alpha|^2 + |\beta|^2)})}} \left( 
|\alpha \rangle |\beta\rangle + |-\alpha\rangle|-\beta\rangle \right).
\end{equation}
Let's set the following orthonormal basis of system A
by superposing nonorthogonal and linear independent two coherent states $|\alpha \rangle $ and $|-\alpha \rangle$ 
\begin{eqnarray}\label{cb1}
|e_1 \rangle &=& \frac{1}{\cos 2\theta}(\cos\theta |\alpha \rangle - \sin\theta |-\alpha \rangle);\\\label{cb2}
| e_2 \rangle &=& \frac{1}{\cos 2\theta} (-\sin\theta |\alpha \rangle + \cos\theta |-\alpha \rangle ), 
\end{eqnarray}
with $\sin 2\theta  = \langle - \alpha | \alpha \rangle =  e^{-2|\alpha|^2}$ and $\langle e_i | e_j \rangle = \delta_{ij},\: i = 1, 2$, and $j= 1, 2$. In the same manner, an orthonormal basis of system B can be set by superposing nonorthogonal and linear independent two coherent states $| \beta \rangle $ and $|-\beta \rangle $ as follows
\begin{eqnarray}\label{cb3}
|f_1 \rangle &=& \frac{1}{\cos 2\theta'}(\cos\theta' |\beta \rangle - \sin\theta' |-\beta\rangle);\\\label{cb4}
| f_2 \rangle &=& \frac{1}{\cos 2\theta'} (-\sin\theta' |\beta \rangle + \cos\theta' |-\beta \rangle ), 
\end{eqnarray}
with $\sin 2\theta'  = \langle - \beta | \alpha \rangle =  e^{-2|\beta|^2}$ and $\langle f_k | f_l \rangle = \delta_{kl},\: k = 1, 2$, and $l= 1, 2$.
In term of orthonormal basis we have 
\begin{eqnarray}
|\alpha \rangle &= & \cos\theta |e_1\rangle + \sin\theta |e_2\rangle \\
|-\alpha \rangle &= & \sin\theta |e_1\rangle + \cos\theta |e_2\rangle
\end{eqnarray}
and 
\begin{eqnarray}
|\beta \rangle &= & \cos\theta' |f_1\rangle + \sin\theta' |f_2\rangle; \\
|-\beta \rangle &= & \sin\theta' |f_1\rangle + \cos\theta' |f_2\rangle.
\end{eqnarray}
With respect to the equations (\ref{cb1}) and (\ref{cb2}),
(\ref{cb3}), (\ref{cb4}),
the quantum channel shared by Amy and Bella  $\Psi_{\alpha,\beta}$ in equation (\ref{abq}} takes the form 
\begin{eqnarray}\label{abqr}\nonumber
 | \Psi \rangle_{e, f} &\equiv &
\frac{1}{\sqrt{2( 1+ \sin 2\theta \sin 2\theta'})} \left(
\cos (\theta -\theta') |e_1\rangle|f_1 \rangle \right.\\
&+& \left. \sin(\theta + \theta') |e_1\rangle |f_2\rangle 
+ \sin (\theta + \theta')|e_2\rangle |f_1 \rangle 
+ \cos (\theta -\theta') |e_2\rangle |f_2 \rangle
\right).
\end{eqnarray}
Amy wants to send to Bella the state 
\begin{equation}
|\psi \rangle_a = \cos\theta|e_1\rangle_a + \sin\theta|e_2\rangle_a
\end{equation}
through the channel (\ref{abqr}).
Amy has now two qubits, the one subscribed \textquotedblleft a \textquotedblright which she wants to teleport and one of the entangled pair labeled \textquotedblleft e \textquotedblright , and Bella has one particle labeled \textquotedblleft f \textquotedblright. 

The state of the three particles before Amy's measurement is given by 
\begin{eqnarray}\nonumber
|\Psi\rangle_{aef} = |\psi\rangle_a \otimes |\Psi\rangle_{ef} & = & \left[(\cos\theta |e_1\rangle_a+
\sin\theta |e_2\rangle_a\right]
 \otimes  \frac{1}{\sqrt{2( 1+ \sin 2\theta \sin 2\theta'})}\left[ \right.\\\nonumber
 &{}& \left. \left(
\cos (\theta -\theta') |e_1\rangle|f_1 \rangle 
 + \sin(\theta + \theta') |e_1\rangle |f_2\rangle 
\right.\right.\\\label{totst}
&{}& \left.\left.
+ \sin (\theta + \theta')|e_2\rangle |f_1 \rangle 
+ \cos (\theta -\theta') |e_2\rangle |f_2 \rangle
\right)\:\: \right].
\end{eqnarray}
 In the equation (\ref{totst}), each direct product 
 $|e_1\rangle_a |e_i\rangle, \:, i = 1, 2$ can be expressed in terms of the quasi-Bell operators basis vectors.
 In analogy with the four Bell states  $|\Phi^+\rangle,\;|\Phi^-\rangle,\; 
|\Psi^+\rangle,\; |\Psi^-\rangle $, we follow the general identities applied to the qubits subscribed with \textquotedblleft a \textquotedblright and  labeled with
\textquotedblleft e \textquotedblright as follows
 \begin{eqnarray}
 |\Phi^+\rangle_{ae} & = & \frac{1}{\sqrt{2}}\left(
 |e_1\rangle_a |e_1\rangle + |e_2\rangle_a|e_2\rangle
 \right);\\
  |\Phi^-\rangle_{ae} & = & \frac{1}{\sqrt{2}}\left(
 |e_1\rangle_a |e_1\rangle - |e_2\rangle_a|e_2\rangle
 \right);\\
  |\Psi^+\rangle_{ae} & = & \frac{1}{\sqrt{2}}\left(
 |e_1\rangle_a |e_2\rangle + |e_2\rangle_a|e_1\rangle
 \right);\\
 |\Psi^+\rangle_{ae} & = & \frac{1}{\sqrt{2}}\left(
 |e_1\rangle_a |e_2\rangle - |e_2\rangle_a|e_1\rangle
 \right).
 \end{eqnarray}
Then 
\begin{eqnarray}\label{belid1}
|e_1\rangle_a |e_1 \rangle &= & \frac{1}{\sqrt{2}}
\left(|\Phi^+\rangle_{ae} + |\Phi^-\rangle_{ae} \right);\\\label{belid2}
|e_1\rangle_a |e_2 \rangle &= & \frac{1}{\sqrt{2}}
\left(|\Psi^+\rangle_{ae} + |\Psi^-\rangle_{ae} \right);\\\label{belid3}
|e_2\rangle_a |e_1 \rangle &= & \frac{1}{\sqrt{2}}
\left(|\Psi^+\rangle_{ea} - |\Psi^-\rangle_{ea} \right);\\\label{belid4}
|e_2\rangle_a |e_2 \rangle &= & \frac{1}{\sqrt{2}}
\left(|\Phi^+\rangle_{ea} - |\Phi^-\rangle_{ea} \right).
\end{eqnarray}
 Applying the identities (\ref{belid1}), (\ref{belid2}), (\ref{belid3}), (\ref{belid4}), and expanding, the equation 
(\ref{totst} becomes:
\begin{eqnarray}\nonumber
|\Psi\rangle_{aef} &= &
\frac{1}{2\sqrt{1 + \sin 2\theta \sin 2\theta'}}
\left\lbrace \right.\\\nonumber
{}\\\nonumber
&{}&\left. |\Phi^+\rangle_{ae}\,\otimes [(\cos\theta\cos(\theta -\theta') +
\sin\theta\sin(\theta + \theta'))|f_1\rangle\right.\\\nonumber
&{}& \left. + \,(\sin\theta\cos(\theta - \theta') + \cos\theta\sin(\theta +\theta')) |f_2\rangle \,]\right.\\\nonumber 
&+&\\\nonumber
&{}& \left. 
|\Phi^-\rangle_{ae}\,\otimes[(\cos\theta\cos(\theta -\theta') -
\sin\theta\sin(\theta + \theta'))|f_1\rangle\right.  \\\nonumber
&{}& \left. + \,(\cos\theta\sin(\theta +\theta') -
\sin\theta\cos(\theta - \theta')) |f_2\rangle \,]\right.\\\nonumber
&{+}& \\\nonumber
 &{}& \left. |\Psi^+\rangle_{ae}\,\otimes[(\cos\theta\sin(\theta + \theta') +
\sin\theta\cos(\theta - \theta'))|f_1\rangle\right.\\\nonumber
&{}& \left. + \,(\cos\theta\cos(\theta -\theta') +
\sin\theta\sin(\theta + \theta')) |f_2\rangle)\,]\right.\\\nonumber
&{+}& \\\nonumber
 &{}& \left. |\Psi^-\rangle_{ae}\,\otimes[(\cos\theta\sin(\theta + \theta') -
\sin\theta\cos(\theta - \theta'))|f_1\rangle\right.\\\nonumber
&{}& \left. + \,(\cos\theta\cos(\theta -\theta') -
\sin\theta\sin(\theta + \theta')) |f_2\rangle)\,]\right.
\\
&{}& \left.\right\rbrace.
\end{eqnarray}

Since no operations have been performed, the three qubits are still in the same total state. The teleportation occurs when Amy measures her two qubits (a and e) in the quasi-Bell basis $|\Phi^+ \rangle_{ae}\,, |\Phi^-\rangle_{ae}\,, |\Psi^+ \rangle_{ae}\,, |\Psi^-\rangle_{ae}$.  Amy's two qubits are now entangled to each other in one of the four quasi-Bell states and the entanglement originally shared between Amy's and Bella's qubits is now broken. Since the states $|\psi \rangle_a$ and $|\psi \rangle_e$ are with Amy, she performs a quasi Bell state measurement on her states and send the measurement result to Bella expending two classical bits.
The result of Amy's measurement tells her which of the four following states the system is in
\begin{equation}
|\Phi^+\rangle_{ae}\,\otimes [(\cos\theta\cos(\theta -\theta') +
\sin\theta\sin(\theta + \theta'))|f_1\rangle
 + \,(\sin\theta\cos(\theta - \theta') + \cos\theta\sin(\theta +\theta')) |f_2\rangle];
\end{equation}
\begin{equation}
|\Phi^-\rangle_{ae}\,\otimes[(\cos\theta\cos(\theta -\theta') - \sin\theta\sin(\theta + \theta'))|f_1\rangle
 + \,(\cos\theta\sin(\theta +\theta') -
\sin\theta\cos(\theta - \theta')) |f_2\rangle \,];
\end{equation}
\begin{equation}
|\Psi^+\rangle_{ae}\,\otimes[(\cos\theta\sin(\theta + \theta') +
\sin\theta\cos(\theta - \theta'))|f_1\rangle
+(\cos\theta\cos(\theta -\theta') +
\sin\theta\sin(\theta + \theta')) |f_2\rangle)];
\end{equation}
\begin{equation}
|\Psi^-\rangle_{ae}\,\otimes[(\cos\theta\sin(\theta + \theta') -
\sin\theta\cos(\theta - \theta'))|f_1\rangle
 + (\cos\theta\cos(\theta -\theta') -
\sin\theta\sin(\theta + \theta')) |f_2\rangle)].
\end{equation}
Bella's qubits takes on one of the four superposition states above and they are unitary images of the state to be teleported. After Bella receive the message from Amy, she guesses which of the four states her qubit is in. Using this information, Bella accordingly chooses one of the unitary transformation $\{ \mathbb{I}, \sigma_x,\, i\sigma_y,\, \sigma_z \}$ to perform her part of the channel. Here $\mathbb{I}$ represents the identity operator and $\sigma_x, \sigma_y, \sigma_z$ are the Pauli operators, and the correspondence between the measurement outcomes and the unitary operations are 
\begin{equation}
|\Phi^+\rangle_{ae} \Rightarrow \mathbb{I};\: 
|\Phi^-\rangle_{ae} \Rightarrow \sigma_z;\:
|\Psi^+ \rangle_{ae} \Rightarrow \sigma_x;\:
|\Psi^-\rangle_{ae} \Rightarrow i\sigma_y\,.
\end{equation}
The teleportation is achieved and in order to measure the efficiency of the teleportation protocol we compute the fidelity of this teleportation as discussed in \cite{popescu, horodecki, hardy}
The teleportation fidelity is given by 
\begin{equation}
F^{\textrm{tel}} = \sum_{i = 1}^4 P_i\,|\langle\psi_a |\chi_i \rangle|^2\,,
\end{equation}
where $P_i = \textrm{Tr}({}_{aef}\langle \Psi |M_i|\Psi\rangle_{aef})$, and $M_i = | \psi_i\rangle \langle \psi_i |$ the measurement operator in the quasi-Bell basis 
$|\psi_i \rangle \in \{ |\Phi^+ \rangle_{ae},\:  |\Phi^-\rangle_{ae},\: |\Psi^+\rangle_{ae},\: |\Psi^-\rangle_{ae}  \}$, and  $|\chi \rangle_i $ is the teleported state corresponding to the ith projective measurement in the quasi-Bell basis to the teleported state, that is nothing than Bella's normalized and corrected outcome given the measurement result $i$. Let's compute first $P_i, i = 1, \ldots 4.$
\begin{eqnarray}
P_1 & = & \textrm{Tr}({}_{aef}\langle \Psi | \Phi^+\rangle_{ae}\langle \Phi^+ | \Psi\rangle_{aef});\\
P_2 & = & \textrm{Tr}({}_{aef}\langle \Psi | \Phi^-\rangle_{ae}\langle \Phi^- | \Psi\rangle_{aef});\\
P_3 & = & \textrm{Tr}({}_{aef}\langle \Psi | \Psi^+\rangle_{ae}\langle \Psi^+ | \Psi\rangle_{aef});\\
P_4 & = & \textrm{Tr}({}_{aef}\langle \Psi | \Psi^-\rangle_{ae}\langle \Psi^- | \Psi\rangle_{aef}).
\end{eqnarray}
\begin{eqnarray}
P_1  = P_3 & = & \frac{1}{4} + \frac{1}{4}\left( \frac{\sin^2(2\theta) + \sin 2\theta \sin 2\theta'}{1 + \sin 2\theta \sin 2\theta'}\right);\\
P_2  = P_4  & = & \frac{1}{4} - \frac{1}{4}\left( \frac{\sin^2(2\theta) + \sin 2\theta \sin 2\theta'}{1 + \sin 2\theta \sin 2\theta'}\right).
\end{eqnarray}
The fidelity of transportation the state $|\psi\rangle_a$ given the channel $ |\Psi \rangle_{ef}$ in equation (\ref{abq} is given by
\begin{equation}\label{fidel}
F^{\textrm{Tel}} = \frac{\cos^2(\theta -\theta') + \sin^2(2\theta)\sin^2(\theta +\theta')}{1 + \sin 2\theta \sin 2\theta'}.
\end{equation}
If $\theta = \theta'$, we have 
\begin{equation}
F^{\textrm{Tel}} = \frac{1 + \sin^4(2\theta)}{1 + \sin^2 2\theta }.
\end{equation}
We determine the minimum assured fidelity (MASFI) which corresponds to the least value of possible fidelity for a given information and can be used as measure of quality of teleportation \cite{prakash1, prakash2}.
The concurrence of the state channel that is 
\begin{equation}
C(|\Psi\rangle_{ef}) = \frac{\cos 2\theta \cos 2\theta'}{ 1 + \sin 2\theta\sin 2\theta'}.
\end{equation}
More details on how to compute the concurrence of an entangled state can be found in \cite{gnonfin}.
The minimum assured fidelity (MASFI) is defined as 
\begin{equation}\label{masfi}
(\textrm{MASFI})_{|\Psi\rangle_{ef}} = \frac{2 C(|\Psi\rangle_{ef})}{1 + C(|\Psi\rangle_{ef})} = 
\frac{2 \cos 2\theta \cos 2\theta'}{1 + \sin 2\theta \sin 2\theta' + \cos 2\theta \cos 2\theta'}.
\end{equation}
If $\theta = \theta'$, then 
\begin{equation}
(\textrm{MASFI})_{|\Psi\rangle_{ef}} = \cos^2 2\theta.
\end{equation}

Let's recall that $\sin (2\theta) = \exp{(-2|\alpha|^2)}$ and $\sin (2\beta) = \exp({-2|\beta|^2})$. As we have already noticed, when $|\alpha| \rightarrow \infty$ and $|\beta| \rightarrow \infty$, the quasi-Bell states as defined in equations (\ref{qb1}), (\ref{qb2}), (\ref{qb3}), (\ref{qb4}) are maximal and this is justified here by the fact that when $|\alpha|\rightarrow \infty,\, |\beta| \rightarrow \infty$, then $\sin (2\theta) \rightarrow 1$ and $\sin (2\theta') \rightarrow 1$, that means that $\theta \rightarrow (\pi/4)$ and $\theta' \rightarrow (\pi/4)$.
For $\theta = \theta' = (\pi/4)$, the fidelity in the equation (\ref{fidel}) is 1 and the MASFI in equation 
(\ref{masfi}) is 0.

\section{Concluding remarks}\label{sec5}
In this work, quasi-Bell states have been established using non-orthogonal states. The discussion about teleportation via one of these quasi-Bell states has been motivated by the Horodecki criterion which has shown that the teleportation scheme obtained by replacing the quantum channel (Bell state) of the usual teleportation scheme with a quasi-Bell state is optimal. Indeed, in a recent work on a comparative study of teleportation of a qubit using entangled non-orthogonal states, it has been established that all the quasi-Bell states, which are entangled non-orthogonal states, may be used for quantum teleportation of a single qubit state \cite{mitali}. In our investigation, we have shown that the performance of the teleportation depends on the maximality of the quasi-Bell states and not on the non-orthogonality of the coherent states. The minimum assured fidelity (MASFI) cannot be unity as for specific cases discussed in \cite{mitali}. At $\theta = \theta' = (\pi/2)$, the minimum assured fidelity (MASFI) $= 1$ ) and the concurrence as well as the fidelity take the value unit, but this is in contradiction with the fact that 
$\sin (2\theta) = \exp{(-2|\alpha|^2})$ due to the non-orthogonality of the coherent states. In conclusion, deterministic perfect teleportation is not possible in the case of our study. We hope the present study will be useful for the investigations of entangled non-orthogonal coherent states.
\section*{Acknowledgments}
The work of IA is supported by the World Bank through CERME (Centre d'Excellence R\'egional pour la Ma\^itrise de l'Electricit\'e ).


\begin{thebibliography}{10}
\addcontentsline{toc}{chapter}{References}
\bibitem{aremua-gouba1}
I. Aremua and L. Gouba, \emph{Coherent states for a system of an electron moving on plane,} {\it J. Phys. Commun.}  {\bf 5}  (2021) 085013.

\bibitem{aremua-gouba2}
I. Aremua and L. Gouba, \emph{Coherent states for a system of an electron moving in a plane: Case of discrete spectrum,} {\it J. Phys. Commun..}  {\bf 5 }  (2021) 125009.

\bibitem{aremua-gouba} 
I. Aremua and L. Gouba, \emph{ Unitary maps on Hamiltonians of an electron moving in a plane and coherent state construction, } {\it J. Math. Phys.} {\bf 64} (2023) 
063508.

\bibitem{gazeauklauder}
J P Gazeau and J. R, Klauder, \emph{Coherent states for systems with discrete and continuous spectrum}, {\it J. Phys. A: Math. Gen.} {\bf 32 } (1999) 123 - 132.

\bibitem{alibagarello}
 S. T. Ali and F. Bagarello, \emph{Some physical appearances of vector coherent states and coherent states related to generate Hamiltonians}, 
 {\it J. Math. Phys.} {\bf 46} (2005) 053518.
 
\bibitem{gazeau-novaes}
J P Gazeau and M. Novaes, \emph{multidimensional generalized coherent states } {\it J. Phys. A: Gen.} {\bf 36} (2003) 199-212.

\bibitem{lgouba}
L. Gouba, \emph{Time dependent q-deformed bi-coherent states for generalized uncertainty relations}, {\it J. Math. Phys.} {\bf 56} (2015) 073507.
 
 \bibitem{peter1}
C. Duval and P. A. Horv\'athy, 
\emph{Exotic Galilean symmetry in the non-commutative plane and the Hall effect } {\it J. Phys. A: Math. Gen.} {\bf 34}  (2001) 10097.

\bibitem{peter2}
P. A. Horv\'athy
\emph{ The Non-commutative Landau Problem }
{\it Annals of Physics} {299}, Issue 1, (2002), 128-140.

\bibitem{peter3}
Peter A. Horv\'athy, Luigi Martina and Peter C. Stichel,
\emph{ Exotic Galilean Symmetry
and Non-Commutative Mechanics } {\it SIGMA} {\bf 6} (2010), 060.

\bibitem{peter4}
P. M. Zhang, P. A. Horv\'athy
\emph{Kohn condition and exotic Newton-Hooke symmetry in the non-commutative Landau problem } {\it Physics Letters B} {\bf 706}, Issues 4-5, (2012), 442-446.

\bibitem{aremua-hounkonnou-balo}
Isiaka Aremua, Norbert Mahouton Hounkonnou, and Ezinvi Balo\"itcha
\emph{ Density operator formulation for magnetic
systems: Physical and mathematical
aspects }, {\it J. Math. Phys.} {\bf 62} (2021) 013503.

\bibitem{sanders1}
Barry C. Sanders 
\emph{Entangled coherent states} 
{\it Phys. Rev. A} {\bf 45}, (1992) 6811. {\it Erratum Phys. Rev. A} {\bf 46}, (1992) 2966.

\bibitem{sanders2}, 
Xiaoguang Wang,  Barry C. Sanders and Shao-hua Pan 
\emph{ J. Phys. A: Math. Gen.} {\bf 33} (2000) 7451.

\bibitem{sanders3}
Barry C. Sanders,
\emph{Review of entangled coherent states } 
{\it J. Phys. A: Math. Theor. } {\bf 45} (2012) 244002.

\bibitem{sanders4} 
Barry C. Sanders,
\emph{ Forty-five years of entangled coherent states  } 
{it Proceedings of the First International Workshop on ECS and Its Application to QIS;T.M.Q.C.} 
( 2013) 111-113.

\bibitem{gnonfin}
H. Gnonfin, L. Gouba
\emph{The Separability Problem in Two Qubits Revisited}
{\it Symmetry} {\bf 15} Issue 11 (2023), 2089.

\bibitem{bennett}
C. H. Bennett, G. Brassard, C. Crepeau, R. Jozsa, A. Peres, W. K. Wootters, \emph{Teleporting an unknown quantum state vis dual classical and Einstein-Podolsky-Rosen channels} {\it Phys. Rev. Lett } {\bf 70} (1993) 1895.

\bibitem{popescu}
S. Popescu 
\emph{Bell's inequalities versus teleportation: What is nonlocality?}
{\it Phys. Rev. Lett.} {\bf 72}  (1994) 797

\bibitem{horodecki}
R. Horodecki, M. Horedecki, P. Horodecki 
\emph{Teleportation, Bell's inequalities and inseparability} {\it  Phys. Lett. A} 
{\bf 222}, Issues 1-2, (1996), 21-25.

\bibitem{hardy}
Henderson, L., Hardy, L., Vedral, V. \emph{Two-state teleportation} {\it Phys. Rev. A} {\bf 61} (2000) 062306.

\bibitem{prakash1}
Prakash, H., Chandra, N., Prakash, R. 
\emph {Improving the teleportation of entangled coherent states} {\it Phys. Rev. A} {\bf 75} (2007) 044305.

\bibitem{prakash2}
Prakash, H., Verma, V. {\it Minimum assured fidelity and minimum average fidelity in quantum teleportation of single qubit using non-maximally entangled states}
{\it Quantum Inf. Process} {\bf 11} (2012) 1951-1959

\bibitem{mitali}
Mitali Sisodia, Vikram Verma, Kishore Thapliyal, Anirban Pathak \emph{Teleportation of a qubit using entangled non-orthogonal states: a comparative study} {\it Quantum Inf Process (2017)} {\bf 16} 76.





\end{thebibliography}
\end{document}